# Evidence for a new extended solid of nitrogen

Li Lei[1,2,*], Qiqi Tang[1], Feng Zhang[1], Shan Liu[1], Binbin Wu[1], Chunyin Zhou[3]

*1, Institute of Atomic and Molecular Physics, Sichuan University, 610065 Chengdu, China*

*2, Key Laboratory of High Energy Density Physics and Technology (Ministry of Education), Sichuan University, 610064, Chengdu, China*

*3, Shanghai Synchrotron Radiation Facility, Shanghai Advanced Research Institute, Chinese Academy of Sciences, 201204 Shanghai, China*

*\*Electronic mail: lei@scu.edu.cn*

**Abstract**

A new extended solid nitrogen, referred to post-layered-polymeric nitrogen (PLP-N), was observed by further heating the layered-polymeric nitrogen (LP-N) to above 2300 K at 161 GPa. The new phase is found to be very transparent and exhibits ultra-large *d*-spacings ranging from 2.8 to 4.9 Å at 172 GPa, suggesting a possible large-unit-cell 2D chain-like or 0D cluster-type structure with wide bandgap. However, the observed X-ray diffraction pattern and Raman scattering data cannot match any predicted structures in the published literatures. This finding further complicates the phase diagram of nitrogen and also highlights the path dependence of the high-pressure dissociative transition in nitrogen. In addition, the forming boundary between cg-N and LP-N has been determined.

The understanding pressure-induced dissociative transition from dense molecular solids to nonmolecular extended solids in low-Z elements, such as hydrogen [1-3] and nitrogen [4-6], is an important objective in high-pressure physics. The high-pressure dissociation and metallization transition in nitrogen behave similarly in many respects to hydrogen [7], however, nitrogen differs from hydrogen in its chemical bonding property forming rich solid polymorphous [8]. The intermolecular dissociation transition in nitrogen involved with rearrangement of triple bonds, resulting in various molecular solids, such as β [9,10], δ [11-13], $δ_{loc}$ [13,14], ε [14,15], ζ [16,17], ζ' [18, 22] κ[17], λ [19,20], η [18], θ [21] and ι [7, 21]. The intramolecular dissociation transition in nitrogen is correlated with the rupture of strong covalent triple bonds, resulting in different densely packed, wide bandgap, single-bonded polymeric phases, such as cg-N [6], LP-N [25] and HLP-N [26]. The resulting polymeric phases typically have high energy density (33 kJ/cm$^3$ for cg-N) [27] and have promise as a high-energy-density material (HEDM).

Direct compression molecular nitrogen to megabar often yields extended solids associated with the pressure-induced symmetry breaking [22]. In spite of considerable theoretical efforts have been made on the prediction of polymeric structures with rings [28], layers [29], clusters [30], chains [31] and cages [32]; there are still considerable disagreements between theoretical predictions [30-32] and experimental observations [26, 28]. More than 20 polymeric phases have been predicted from theoretical calculations [28-33], but only three types of polymeric phases were experimentally observed so far [6, 22-26]. There are still controversies over the issue of the structure for the polymeric phases LP-N [26, 28], and the *true* crystal structure of LP-N requires unambiguous determination. Strong diffraction spots rather than powder rings often present on the high-pressure X-ray diffraction pattern due to preferred spontaneous crystallization at high pressure and high temperature, and the absent Raman scattering often occurs in high-pressure Raman spectroscopy because of the extreme strain environment in the small sample cavity of DAC. Therefore, it has been challenging to achieve a convincing result for the structure of extended solid nitrogen.

The complexity of phase relation in nitrogen is partially attributed to the path-dependence transformation under high pressure. The observation of the thermodynamically favorable high-pressure phase depends on the experimental pressure-temperature (*P-T*) path we selected, so walking different *P-T* paths may allow us access to a different metastable phase, which would provide a better understanding of pressure-induced dissociative transition in nitrogen. Here, we loaded the solid nitrogen in two different *P-T* pathways (a red path and a blue path), and presented the experimental observation of a novel high-pressure extended nitrogen, post-layered-polymeric nitrogen (PLP-N), beyond the *P-T* stability field of LP-N. However, the observed X-ray diffraction pattern and Raman scatting data for the PLP-N cannot match any predicted structure of nitrogen in the literatures.

High pressure was generated using a diamond-anvil cell (DAC) with 80~100 μm culets. Rhenium was used as the gasket material precompressed to 20 μm thickness with a sample chamber drilled using laser cutting to produce a 25 μm diameter holes. The pressure was monitored by the high-frequency edge of the diamond phonon [34].

High-pressure Raman experiments were carried out on a custom-built confocal Raman spectrometry system in the back-scattering geometry based on Andor Shamrock triple grating monochromator with an attached Andor Newton EMCCD, excitation by a solid-state laser at 532 nm. The high-pressure angle dispersive X-ray diffraction (ADXRD) data were collected at the BL15U1 beamline of the Shanghai Synchrotron Radiation Facility (SSRF, China). A monochromatic X-ray beam with an energy of 20 keV and a focused beam size of 2.9×3.2 μm$^2$ was used for ADXRD measurements. The two *P-T* paths are shown in Fig. 1. Laser heating experiments were conducted without the use of irradiation absorbing agents in a custom-build double-sides laser-heating DAC system with two 1064 nm CW fiber lasers with ~5 μm laser-heating spot size. Laser heating temperature was either measured by spectroradiometric method or estimated [22]. About 3-5 GPa pressures were dropped at the transition point after laser heating due to the structural relaxation and the relieving of stress. The pressure release was stopped due to a diamond anvil failure at ~70 GPa, and no apparent evidence was found in the spectral and diffraction data, suggesting chemical reaction of nitrogen with diamond or rhenium gasket.

In the red path, the representative Raman spectra are shown in Fig. 2a, upon compression to above 130 GPa, the molecular nitrogen becomes dull red and Raman silent. the amorphous "red"-N formed at ~1000 K and 131 GPa can well absorb near-infrared laser radiation (1064 nm) and be laser heated directly without the addition of a thermal absorbing agent. After direct laser-heated the "red"-N to above 1800 K at 134 GPa, the cg-N is formed as evidenced by the observation of the fingerprint Raman *A* mode of cg-N at ca. 855 cm$^{-1}$. Upon further room-temperature compression to 157 GPa, the *A* mode of the cg-N becomes very weak. But the cg-N transforms to the LP-N after successive heating up to a~2000 K as evidenced by the presence of Raman modes of the LP-N at ca. 857 cm$^{-1}$, 1033 cm$^{-1}$, and 1150 cm$^{-1}$. It is shown that the forming region of LP-N is above the *P-T* stability field of the cg-N, and the phase boundary between cg-N and LP-N can be determined at 157 GPa and ~2000 K (Fig. 1). It is also noted that the main Raman peak at 1033 cm$^{-1}$ of the LP-N exhibits colossal Raman scattering that is first reported by Tomasino, *et al.* [25]. Upon further room-temperature compression to 190 GPa, the Raman peaks of LP-N exhibits blue shift and significant broadening, indicating wide pressure stability over the pressure range from 157 GPa to 190 GPa. The above-described case is the red *P-T* path as shown in the Fig. 1.

In the blue *P-T* path, the representative Raman spectra are shown in Fig. 2b, the cg-N is initially formed under lower pressure and higher temperature condition (114 GPa and ~2000 K) as compared with the red path. After successive laser heating to above 2200 K at 131 GPa and 151 GPa, the cg-N is still unable to transform into the LP-N due to the lack of pressure. Upon laser heating at a higher pressure (159 GPa), however, a mixed-phase of cg-N and LP-N is produced, which further confirmed the phase boundary (157 GPa, 2000 K) between the cg-N and LP-N (Fig. 1). We also noted that the intensity of Raman peak of the obtained LP-N is very low as compared with the colossal Raman scattering of the LP-N sample in the red path, suggesting the existence of a metastable state far away from thermodynamic equilibrium in the sample. We continuedly compressed the mixed-phase sample to 161 GPa, and further laser-heated the sample to a higher temperature (~ 2300 K) for the purpose of producing phase-pure LP-N. Unexpectedly, all the Raman bands for the cg-N and LP-N are found to be disappeared, but three Raman modes at 985 cm$^{-1}$, 1021 cm$^{-1}$, and 1054 cm$^{-1}$ at 161 GPa along with strong luminescence background are observed. After the laser-heating at 161 GPa, the sample now becomes very transparent, more transparent than the cg-N (Fig. 1 inset), suggesting a higher bandgap value of the new state. This result is most probably associated with the formation of a novel extended solid phase that is different from the previously reported structures, including cg-N [6], LP-N [25] and HLP-N [26]. Because the new phase was produced by further heating the LP-N to 2300 K at 161 GPa, beyond the *P-T* stability field of LP-N, the newly synthesized extended solid nitrogen is referred to the post-layered-polymeric nitrogen, PLP-N. This result further complicates the phase diagram of nitrogen. Recent fast optical spectroscopy experiments to just higher temperatures than the transition point suggested a metallization transition in nitrogen [7]. The observation of wide-bandgap PLP-N, however, cannot provide direct evidence of the insulator-to-metal transformation. The bandgap of the crystalline polymeric nitrogen does not decrease appreciably with pressure.

The Raman peaks of the transparent PLP-N becomes indistinguishable from the background upon further compression to 172 GPa (Fig. 2b), even after repetitive laser heating at this pressure. Noted that there are at least five Raman modes at 700-1100 cm$^{-1}$ observed on the decompression. Phonon splitting implies the pressure-induced symmetry lowering transition from 3D network cg-N to 2D layered LP-N, and then to PLP-N with lower symmetry (2D or 0D). The three Raman modes of the PLP-N (the modes 'a', 'b', 'c' in the Fig. 2b and Fig. 3) are analogous to the main Raman peak of LP-N at 1025 cm$^{-1}$ at 150 GPa. The observed Raman mode 'd' exhibits different pressure dependence from the *A* mode of cg-N, especially at pressures lower than 110 GPa. It cannot be rule out the possibility of the reversible transformation into the cg-N on the decompression. As shown in Fig. 3, the pressure-dependent phonon frequency changes in PLP-N suggest a stiff lattice and shift nearly linearly with pressure at approximately 1.1 cm$^{-1}$/GPa. This value is smaller than those of cg-N (~1.3 cm$^{-1}$/GPa) [6, 22] and LP-N (1.2-1.6 cm$^{-1}$/GPa) [25-26], suggesting the PLP-N might have a higher density than that of the cg-N (4.5 g/cm$^3$ at 120 GPa) and LP-N (~4.85 g/cm$^3$). In addition, a broad Raman band at ca. 900 cm$^{-1}$ is also observed in the high-pressure Raman spectra (Fig. 2b), which could come from stress-induced disordered local phonon of nitrogen.

In order to probe the structure of the newly synthesized PLP-N, *in-situ* high-pressure X-ray diffraction experiments have been conducted. As shown in Fig. 4, the new phase exhibits a distinctive X-ray diffraction pattern. The diffraction-peaks positions of the PLP-N are listed in Table I. The X-ray diffraction patterns shown were collected from the center of the sample cavity (~10 μm in size), there are at least eight reflections of the PLP-N. If we focus the X-ray diffraction spot onto the edge of the sample cavity, strong reflections from the Rhenium gasket are found to be observed at the smaller *d*-spacings except for the eight diffraction peaks of PLP-N. We cannot observe the reflections form the cg-N and LP-N. We would like to note that the PLP-N exhibits ultra-large *d*-spacings ranging from 2.8 to 4.9 Å at 172 GPa, indicating the larger primitive cell dimensions that are comparable to other previously known polymeric phases, in which the overwhelming majority of *d*-spacings are smaller than 3.0 Å. The structure of PLP-N is most likely the 2D chain-like or 0D cluster-type structure. We try to fit the measured spectra to the predicted structures; however, the observed X-ray diffraction pattern and Raman scatting data cannot match any structure in the published literature. Although the exact structure of the new phase is still an open issue, it suggests the wide bandgap and large-unit cell structure properties for the extended solid nitrogen beyond the stability field of cg-N and LP-N. This finding further complicates the phase diagram of nitrogen and also highlights the path dependence of the high-pressure dissociative transition in nitrogen.

In conclusion, we investigate the structural transition of extended solid nitrogen via two different *P-T* paths. The phase boundary between the cg-N and LP-N has been determined, and a novel high-pressure extended solid, PLP-N, formed above the *P-T* stability field of LP-N is found to be very transparent and is most likely to have a 2D chain-like or 0D cluster-type structure.

We thank Jian Sun for helpful discussion. We acknowledge support by the National Natural Science Foundation of China (Grant No. 11774247) and the Chinese Academy of Sciences (Grant No. 2019-SSRF-PT-009588).

**Table I.** The diffraction-peak positions of PLP-N at 123 GPa and 172 GPa ($\lambda= 0.6199$Å).

| Pressure (GPa) | Peak labels in Fig. 4a | d-spacing (Å) | 2θ (degree) | Relative Intensity |
|---|---|---|---|---|
| 172 | $P_1$ | 2.8432 | 12.4934 | 3 |
| | $P_2$ | 3.0081 | 11.8087 | 26 |
| | $P_3$ | 3.1332 | 11.3373 | 18 |
| | $P_4$ | 3.2010 | 11.0972 | 53 |
| | $P_5$ | 3.9427 | 9.0102 | 68 |
| | $P_6$ | 4.3567 | 8.1544 | 95 |
| | $P_7$ | 4.4315 | 8.0168 | 100 |
| | $P_8$ | 4.8678 | 7.2986 | 43 |
| 123 | $P_1$ | 2.8697 | 12.3781 | 2 |
| | $P_2$ | 3.0401 | 11.6844 | 6 |
| | $P_3$ | 3.1977 | 11.1087 | 5 |
| | $P_4$ | 3.2370 | 10.9739 | 11 |
| | $P_5$ | 4.0088 | 8.8617 | 100 |
| | $P_6$ | 4.3926 | 8.0878 | 20 |
| | $P_7$ | 4.4722 | 7.9439 | 26 |
| | $P_8$ | 4.9094 | 7.2368 | 51 |

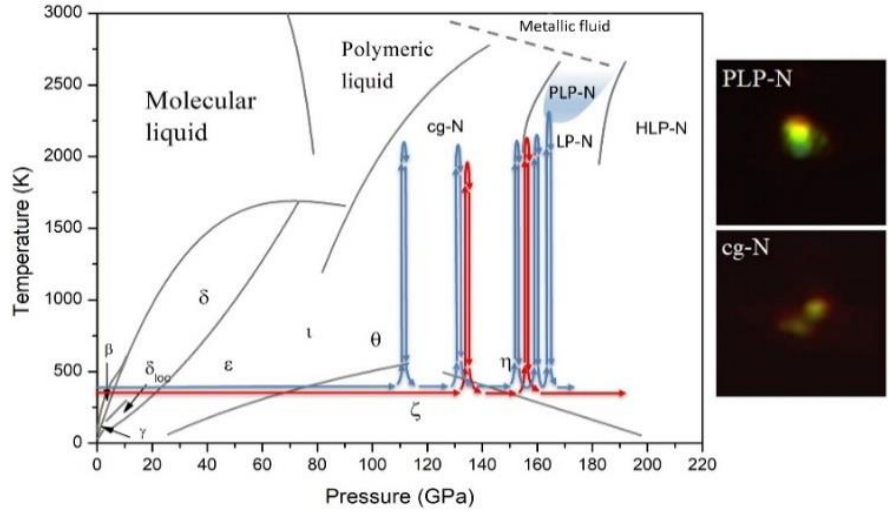

**FIG. 1.** *P-T* phase diagram of nitrogen and two experimental *P-T* routes. Gray solid and dashed lines are the phase boundaries, red and blue arrow lines indicate the two *P-T* paths, blue zone shows the forming region of PLP-N. Insets are microscope images of cg-N at 130 GPa and PLP-N at 172 GPa with the same transmitting light, the sample size is ~10 μm.

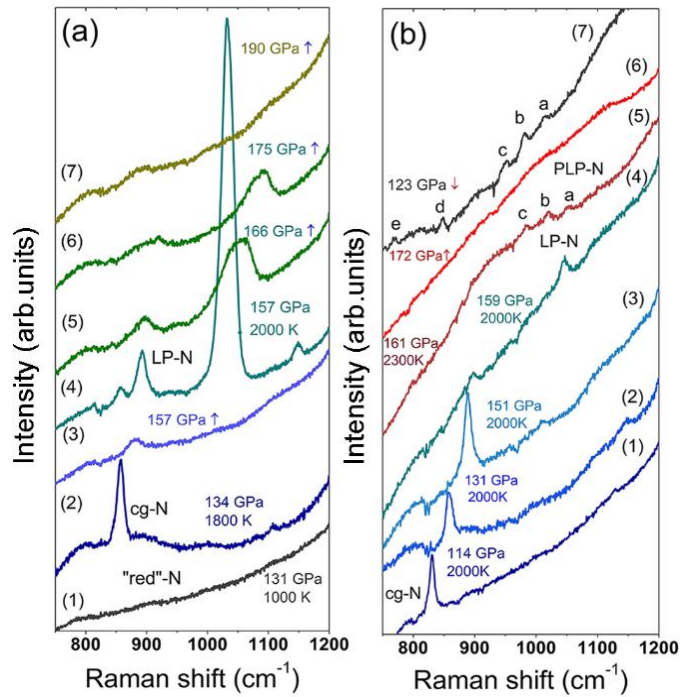

**FIG. 2.** Representative Raman spectra of nitrogen via the red path (a) and the blue path (b). The "↑" and "↓" arrows denote the room-temperature compression and decompression, respectively.

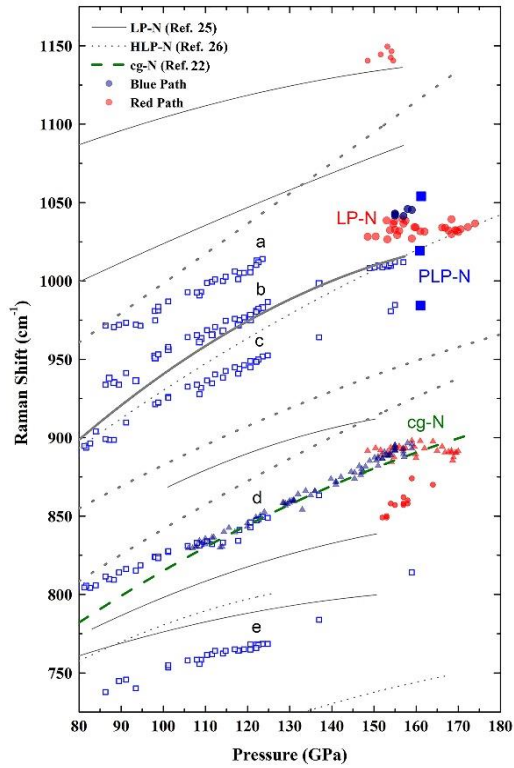

**FIG. 3.** Pressure dependent Raman shifts of PLP-N (square) shown in comparison with those of cg-N (triangle) and LP-N (circle). Red and blue color symbols represent Raman data in the red path and the blue path, respectively. Solid and open symbols represent the present Raman data on compression and decompression, respectively. Dashed lines, solid lines and dotted lines are the previously published data for cg-N [22], LP-N [25] and HLP-N [26]. Thick lines and larger symbols indicate the main peaks of Raman spectra.

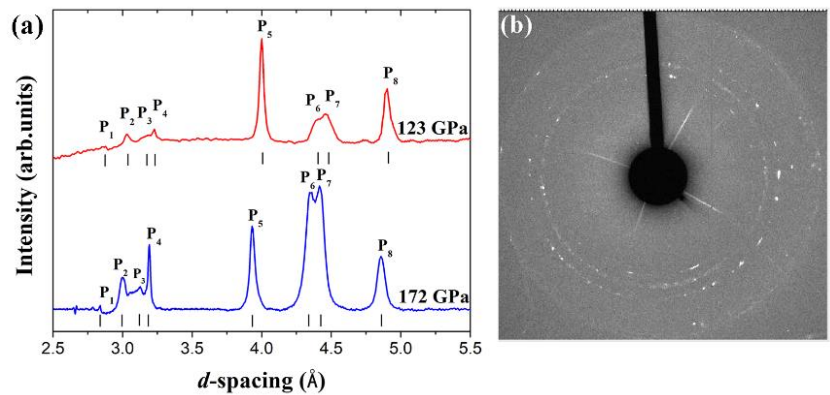

**FIG. 4.** (a) The integrated X-ray diffraction patterns of PLP-N taken at room temperature at 123 GPa and 172 GPa from the center of the nitrogen sample, black tick lines indicate the positions for the observed diffraction $d$-spacing. (b) A typical 2D X-ray diffraction image of the PLP-N at 123 GPa.